\def\year{2020}\relax
\documentclass[letterpaper]{article} 
\usepackage{aaai20}  
\usepackage{times}  
\usepackage{helvet} 
\usepackage{courier}  
\usepackage[hyphens]{url}  
\usepackage{graphicx} 
\usepackage{graphicx}  
\usepackage{times}
\usepackage{subfigure}
\usepackage{soul}
\usepackage[utf8]{inputenc}
\usepackage[small]{caption}
\usepackage{graphicx}
\usepackage{amsmath}
\usepackage{amsfonts}
\usepackage{booktabs}
\usepackage{algorithm}
\usepackage{algorithmic}
\usepackage{comment}
\usepackage{url}
\usepackage{scalefnt}
\usepackage{bm}
\usepackage{multirow}
\usepackage{CJKutf8}
\usepackage{enumerate}

\nocopyright

\title{Universal Fake News Collection System using Debunking Tweets}
\author{Taichi Murayama, Shoko Wakamiya, Eiji Aramaki\\
Nara Institute of Science and Technology\\
\{murayama, wakamiya, aramaki\}@is.naist.jp}

\begin{document}
\maketitle

\begin{abstract}
Large numbers of people use Social Networking Services (SNS) for easy access to various news, but they have more opportunities to obtain and share ``fake news'' carrying false information.
Partially to combat fake news, several fact-checking sites such as Snopes and PolitiFact have been founded.
Nevertheless, these sites rely on time-consuming and labor-intensive tasks.
Moreover, their available languages are not extensive.
To address these difficulties, we propose a new fake news collection system based on rule-based (unsupervised) frameworks that can be extended easily for various languages.
The system collects news with high probability of being fake by debunking tweets by users and presents event clusters gathering higher attention.
Our system currently functions in two languages: English and Japanese. 
It shows event clusters, 65\% of which are actually fake.
In future studies, it will be applied to other languages and will be published with a large fake news dataset.
\end{abstract}

\section{Introduction}
Social networking services (SNS) such as Facebook and Twitter have been used widely throughout the world because people can easily and immediately obtain various news and information free of charge.
According to Pew Research Center, 62\% of adults in the United States had received news from SNS in 2017~\cite{pew}.
People continue to benefit from the convenience of excellent sources using SNS, but they have increasing vulnerability to obtaining and sharing news that has not been fact-checked carefully and which includes false or uncertain information, called as ``fake news.''
Fake news is ``a news article or message published and propagated through media, carrying false information regardless the means and motives behind it~\cite{fake_survey2}.''
Some organizations and individuals spreading fake news for financial and political gains cause harm to society.
For example, during the US 2016 presidential election, various tweets related to fake news had been shared more than 37 million times on SNS and had no small effect on the election result~\cite{2016election1,2016election2}.
But it affects not only elections: fake news appears in relation to various events~\cite{disaster1,disaster2,shooting}.

\begin{figure*}[t!]
    \includegraphics[width=17cm]{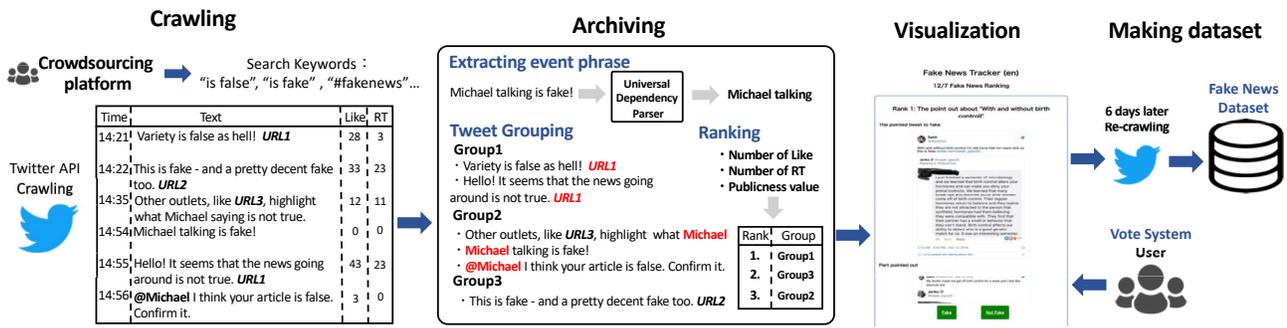}
    \caption{Framework of the proposed system: The system can be divided into four steps: crawling, archiving, visualization, and making dataset.
    The crawling step collects \emph{``fake''}-related tweets.
    The archiving step organizes the collected tweets. Then, the visualization step shows ranked results.
    The making dataset step recrawls tweets to create fake news datasets.}
    \label{overview}
\end{figure*}

Recently, there are various fact-checking websites by domain-experts such as Snopes.com, PolitiFact.com, Factcheck.org and so on.
Also, online tools for tracking fake news on SNS have been developed for various studies and datasets \cite{hoaxy,fakenewstracker}.
Existing online tracking tools collect true and fake news that has been manually annotated or reported by such fact-checking websites.
Although these tracking tools play a crucially important role in the gathering of fake news, they present two major difficulties.
The fact-checking websites contributing these tools, are burdened by time-consuming and labor-intensive tasks.
Because various fake information in SNS spreads rapidly and widely, it is necessary to detect the spread at an earlier stage.
Also, some countries (mainly the US and Europe) have reliable fact-checking websites that provide information related to fake news that can be tracked by the tools.
Therefore, it is difficult to apply existing tracking tools for most countries, including Japan, where no fact-checking organization exists and which use languages other than English, even though many instances of fake news have been detected on SNS in these countries.

To solve these difficulties, we present a new tracking system. 
It requires neither human-annotation nor fact-checking websites to identify spreading fake news quickly in any country.
Our system is based on an assumption that SNS user comments such as \emph{``This is a fake news.''} constitutes a cost-free and real-time clue to catch fake news.
The major features of the proposed system are the following.
\begin{itemize}
    \item The system collects news with high probability of being fake by debunking tweets by Twitter users, not with fake annotation by domain experts and fact-checking websites.
    \item Our current system works in two languages: English and Japanese.
    The system uses a rule-based (unsupervised) method. It can be extended easily for various languages.
    In the future, the system will publish a big multilingual fake news dataset.
    \item Whereas existing systems visualize fake news diffusion for researchers, our system presents diffused fake news contents for the public in real time.
\end{itemize}

\section{Related work}
\subsection{Fake checking websites}
In attempts to combat fake news, various fact-checking websites and organizations have been founded.
PolitiFact\footnote{https://www.politifact.com/} is an independent, non-partisan site for online fact-checking, mainly of U.S. political news and politicians' statements.
Snopes\footnote{https://www.snopes.com/}, one of the first online fact-checking websites, handles political, and other social and topical issues.
Gossipcop\footnote{https://www.gossipcop.com/} investigates fake news in U.S. entertainment stories published in magazines and web news.
Although these fact-checking sites have high reliability, they require time-consuming processes and have poor scalability.

\subsection{Fake tracking tools}
Because fake news is diffused in SNS, it is important to track fake news movements when immediately confirming whether news is fake or investigating the nature of fake news.
To meet these demands, some tracking tools have been announced in some papers.
Hoaxy~\cite{hoaxy} is a framework for collecting and tracking fact-checking information and misinformation related to them.
Users can search for topics in which they are interested and check the diffusion visualization of the respective topics.
FakeNewsTracker~\cite{fakenewstracker} is a system for fake news data collection, detection, and visualization on SNS.
They first collect a fake news source from fake-checking websites.
NewsVerify~\cite{vertify}, a real-time news certification system, starts to track news after user inputs and detects the credibility of events.
~\cite{enquiring} is similar to ours in terms of using tweets including particular phrases.
Although it collects rumors using enquiry phrases such as \emph{``Really?''}, we collect fake news using phrases related to debunking or corrections such as \emph{``This is fake.''}
Additionally, our system is based on a systematic framework that can accommodate identification of multilingual fake news.

\subsection{Fake news datasets produced by tracking tools}
The research community has produced various datasets for fake news detection or similar objectives.
For producing datasets, some fake news trackers are used effectively.
Hoaxy dataset~\cite{hoaxy_data} has been accumulated using Hoaxy.
It consists of retweeted messages with links to either fact-checking or misinformation articles.
FakeNewsNet~\cite{fakenewsnet}, constructed using FakeNewsTracker, contains various information such as news contents, and spatio-temporal and social contexts.

\section{Universal fake news collection system}
We first present an overview of the proposed system.
Then, we introduce details of the respective components in our system.
This system will be presented publicly on the web in two languages: English and Japanese.

\subsection{Overview}
The proposed system has four steps: crawling, archiving, visualization, and making dataset.
Figure~\ref{overview} presents an overall picture of the system framework.
Crawling accumulates tweets that point out \emph{``fake''} or similar tweets.
Archiving organizes the collected data and ranks the data for visualization.
Visualization shows tweets in order corresponding to the degree of attention they receive.
The system provides a voting function from users on whether a tweet is related to fake news or not.
Making dataset recrawls tweets with keywords such as URLs obtained during archiving for producing multilingual and large datasets of fake news on SNS.

\subsection{Crawling}
Debunking patterns as search keywords must be found before collecting debunking tweets.
To find useful patterns, we use crowdsourcing platforms: Amazon Mechanical Turk\footnote{https://www.mturk.com/} for English and Yahoo! Crowdsourcing\footnote{https://crowdsourcing.yahoo.co.jp/} for Japanese.
We ask questions such as \emph{``Write what you would write on an SNS such as Twitter for correction when you find false information (for example fake news.)''}
Then we collect 1,000 answer texts from target language speakers.
To acquire useful debunking patterns, we extract uni/bi/tri/4-gram from answer texts and select high-frequency patterns.
From these high-frequency patterns, human experts further selected those which are independent of any particular fake news.
The patterns we selected are presented in Table~\ref{pattern}.
We use the Twitter Search API to crawl tweets including those patterns.
The crawling is executed continuously and the collected tweets are saved in our database.

\begin{CJK}{UTF8}{ipxm}
\begin{table}[!t]
  \centering
    \caption{Selected debunking patterns for crawling tweets in English and Japanese}
\begin{tabular}{|r|r|} \toprule
    \multirow{5}{*}{English} & (isn't\textbar is not) true\\
    & is (completely) (false\textbar fake)\\
    & Don’t believe everything\\
    & spreading (false\textbar fake)\\
    & \#fakenews\\ \midrule
    \multirow{5}{*}{Japanese} & は(デマ\textbar フェイク)\\
     & (デマ\textbar フェイク\textbar フェイクニュース)です\\
     & (フェイク\textbar 間違い\textbar デマ)である\\
     & というデマ \\
     & (信じ\textbar 拡散し)ない \\ \bottomrule
\end{tabular}
 \label{pattern}
\end{table}
\end{CJK}

\subsection{Archiving}
We organize and rank the crawling data for ease of checking.
This step in turn has three steps: extracting event phrase, tweet grouping, and ranking.
Processing all collected tweets is time-consuming. 
Therefore, we use only tweets, the number shares of which are more than three.
These steps are applied every day on one-day of tweets.

\subsubsection{Extracting event phrase}
This step extracts suspicious event phrases pointed out in debunking tweets.
The extracted event phrases are used for the next step, Tweet grouping, and are also used as headlines for visualization.
For example, \emph{``Michael talking''} is extracted as a suspicious event phrase from the tweet \emph{``Michael talking is fake!''}
We execute no machine learning-based extraction but use rule-based extraction, which can be expanded easily in multiple languages.

For suspicious event phrase extraction handling multiple languages, we use the result of Universal Dependencies (UD)~\cite{universal}, which was developed for collection of treebanks with homogeneous syntactic dependency annotation for various languages.
Actually, UD enables application of the same rules to multiple languages for extraction, with a little adjustment.
We obtained treebank from universaldependencies.org\footnote{https://universaldependencies.org/} and applied a UD parser to each tweet.
In this system, a human expert sets extraction rules, which are shown below:
\begin{enumerate}
    \item Parse a sentence including a debunking pattern in Table~\ref{pattern} based on universal dependencies.
    We designate the debunking pattern as ``\textbf{fake part}'' for these processes.
    \item Extract an event phrase from the sentence based on the following two rules.
    One is that the event phrase has dependency arcs to the fake part, which behave as ``nsubj,'' ``nsubjpass,'' ``dobj,'' ``iobj,'' ``csubj,'' or ``appos.''
    These dependency patterns indicate that it is used grammatically by the objective case, the nominative case, the subject itself as clause, and so on.
    Second is that the location of the event phrase is in advance of that of the fake part.
    However, we do not extract the phrase when it is a demonstrative pronoun such as ``this'' and ``it.''
    \item When we do not find the phrase following rules presented in 2, we set the word; the part depends on which in dependency structure, as ``fake part'' and perform the process 3.
    \item When the fake part is ROOT in 3, we change the sentence including the fake part.
    We set following sentence in English, and the preceding sentence in Japanese as the sentence.
    We perform process 2 after we set the ROOT in the sentence as the fake part.
\end{enumerate}

\subsubsection{Tweet grouping}
This step is designed to gather tweets referring to the same event cluster in the same group.
It is difficult to apply machine-learning-based methods for grouping because kinds of tweets are variable every day.
We then execute a simple and robust rule-based grouping method using the extracted suspicious event phrases and other features such as URL.
The rules of grouping are presented below:
\begin{enumerate}
    \item Set tweets with the same URL into the same group
    \item Set tweets replying to the same tweet into the same group
    \item Calculate the distance between extracting event phrases of each tweet in the above step, using the word mover's distance (WMD)~\cite{wmd}.
    Set tweets that have fewer than threshold $\tau$ into the same group
\end{enumerate}
To calculate WMD, we use word vectors from~\cite{wordvec}.
The threshold $\tau$ was set as 0.25.

\subsubsection{Ranking}
This step ranks each group generated from the above step in order of high attention.
Our ranking method is inspired by an unsupervised method of ~\cite{simplification}.
The method ranks each group according to several features, which are considered to express attention.
We then calculate the average rank over all calculated ranks of features as ranking.
Our system uses three features: \textbf{Number of Like}, \textbf{Number of Retweet} and \textbf{Public score}.
Public score calculates the percentage of followers among Retweet users
The larger the first two features are, the higher the degree of attention becomes.
The smaller the public score becomes, the higher the degree of attention becomes.
When the tweet also spreads to other user than followers, it is more important.
The tweet with the highest attention rank is selected from each group for ranking.

\subsection{Visualization}
\begin{figure}[t!]
    \center
    \includegraphics[width=7.5cm]{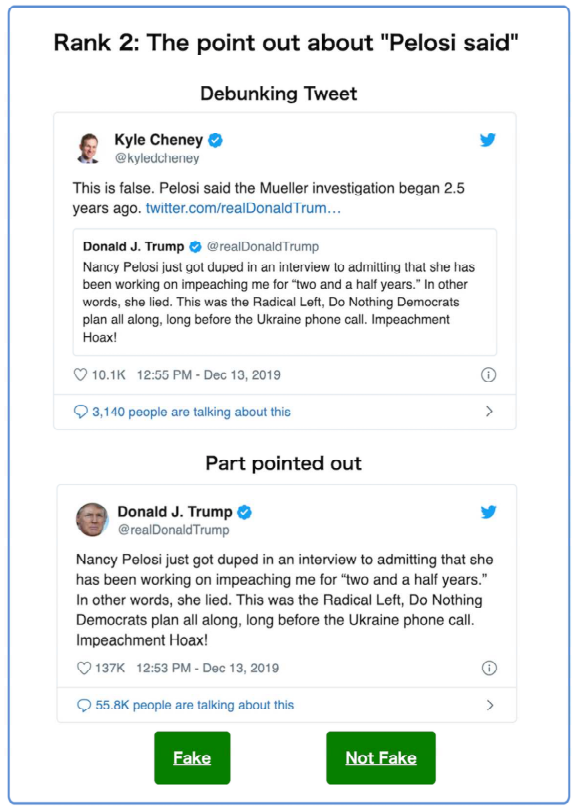}
    \caption{Example of visualization of English fake news.}
    \label{vis}
\end{figure}

The proposed system presents the event clusters in order by our ranking method to meet general demand, not only the researchers'.
An example is presented in Figure~\ref{vis}.

The top 10 event clusters are exhibited in the proposed system. 
Each has three parts: ``Headline,'' ``Debunking Tweet,'' and ``Part pointed out.''
``Headline'' describes suspicious event phrases using ``extracting event phrase.''
``Debunking Tweet'' shows the tweet which the system has obtained from crawling. 
``Part pointed out'' shows the URL, quote tweet, and reply to tweet included in the Debunking Tweet.
The system collects event clusters with a strong probability of being fake, without a fact checking site.

Additionally, we introduce a ``Voting system,'' which enables users to vote on whether each event cluster is fake, or not.
The system clearly shows each event cluster with a strong possibility of being fake by this structure.

\subsection{Making dataset}
We will publish multilingual and large fake news datasets in the future.
We execute re-crawl to event clusters visualized in the system using URL included in the related tweets and keywords obtained by extracting event phrases for producing an exhaustive dataset.
A dataset composed of tweets re-crawled and labeled by the voting system will be published.

\section{Discussion}

\subsection{Representative values of our system}

\begin{table}[!t]
  \centering
  \caption{Representative values of collected tweets and event clusters show daily average numbers of the respective items.}
\begin{tabular}{r|cc} \toprule
     & EN & JA\\ \midrule
    Avg. no. of tweets & 9039 & 7901\\
    Avg. no. of event clusters & 455 & 143\\
    Avg. no. of RT (top10 events) & 1549 & 810\\
    Avg. no. of Like (top10 events) & 5027 & 1899\\
    Avg. no. of verified account (top10 events)  & 2.11 &0.20\\
   \bottomrule
\end{tabular}
 \label{sum}
\end{table}

The system collects numerous tweets daily by continuous crawling.
Table~\ref{sum} shows representative values of tweets and event clusters collected in two languages, English (EN) and Japanese (JA), during November 14, 2019 through December 13, 2019.

No great difference exists between English and Japanese in the numbers of the collected tweets.
By the contrast, event clusters grouped in English are more than three times more numerous than those in Japanese.
This is attributable to the fact that the greater part of the collected tweets in Japanese are retweets.
The numbers of RT and Like of the top event clusters in English are also more than those in Japanese.
This result derives from the situation in which debunked or corrected statements by verified accounts, which have many followers, frequently occur in the top event clusters in English.
Debunked or corrected statements by verified accounts were not found in Japanese.

\subsection{Effectiveness of our system}
Confirming whether a collected event cluster is fake or not is important to validate the effectiveness of the proposed system.
We annotated 124 Debunking tweets (62 tweets in English and 62 tweets in Japanese) visualized from December 7, 2019 to December 13, 2019, for the following viewpoints.
\begin{enumerate}[(a).]
    \item Do sentences in collected tweets indicate debunking?
    \item Are the subjects of collected tweets truly fake?
\end{enumerate}
We recruited two human annotators to label collected tweets manually. We developed a codebook according to the definition of fake news discussed in the Introduction.
Results confirmed a substantial level of agreement: Cohen's Kappa score was 0.73.
For the tweets the two annotators did not agree on, a third annotator (one of the authors) labeled the tweet.

The results of (a) indicates that more than 65\% collected tweets show debunking in each language: 66\% in English and 69\% in Japanese.
This result suggests that selected patterns in Table~\ref{pattern} are appropriate for the system.
The result of (b) also indicates that more than 65\% of subjects of collected tweets are truly fake in each language: 68\% in English and 65\% in Japanese.
The same architecture, irrespective of language, collects event clusters with high probability of being fake.
From these results, we infer that the proposed system achieves a sufficient level of usefulness for practical use.

\section{Conclusions}
Our paper presents a proposal for a fake news collection system to examine debunking tweets specifically.
The system works in two languages: English and Japanese.
By virtue of the fact that the proposed system can be easily extended to other languages, future studies will be undertaken for its application to languages other than English and Japanese and for publishing of a large fake news dataset.
Using the system to gather various fake news items is also expected to contribute to easy comparison of fake news among languages and among countries.

\section{Acknowledgments}
This research was partly supported by Health and Labor Sciences Research Grant Number H30-shinkougyousei-shitei-004 and JSPS KAKENHI Grant Numbers JP19K20279 and JP19H04221.

\bibliographystyle{aaai}
\bibliography{ref}
\end{document}